\documentstyle[11pt,epsfig,amssymb,citesort,axodraw]{article}
\newcommand{\beq}{\begin{eqnarray}}
\newcommand{\eeq}{\end{eqnarray}}
\newcommand{\np}{Nucl. Phys.\ }

\newcommand{\MOM}{\widetilde{\rm MOM}}

\newcommand{\Lams}{\Lambda_{\overline{\rm MS}}}

\newcommand{\Mev}{{\rm MeV}}
\newcommand{\be}{\begin{equation}}
\newcommand{\ee}{\end{equation}}
\newcommand{\lwrsim}{\raise0.3ex\hbox{$<$\kern-0.75em\raise-1.1ex\hbox{$\sim$}}}
\newcommand{\vthree}{\begin{picture}(300,100)(0,0)
\Gluon(50,50)(100,50){5}{4}
\Gluon(100,50)(150,50){5}{4}
\Gluon(150,50)(200,50){5}{4}
\Gluon(200,50)(250,50){5}{4}
\Vertex(100,50){2}
\Vertex(150,50){2}
\Vertex(200,50){2}
\Gluon(100,50)(100,125){5}{3}
\Gluon(150,50)(150,125){5}{3}
\Gluon(200,50)(200,125){5}{3}
\SetWidth{1.4}
\Line(105,130)(95,120)
\Line(105,120)(95,130)
\Line(205,130)(195,120)
\Line(205,120)(195,130)
\end{picture}}
\newcommand{\vfour}{\begin{picture}(300,100)(0,0)
\Gluon(50,50)(125,50){5}{6}
\Gluon(125,50)(200,50){5}{6}
\Gluon(200,50)(250,50){5}{4}
\Vertex(125,50){2}
\Vertex(200,50){2}
\Gluon(125,50)(87.5,125){5}{3}
\Gluon(125,50)(162.5,125){5}{3}
\Gluon(200,50)(200,125){5}{3}
\SetWidth{1.4}
\Line(92.5,130)(82.5,120)
\Line(92.5,120)(82.5,130)
\Line(205,130)(195,120)
\Line(205,120)(195,130)
\end{picture}}

\newcommand{\anthree}{\begin{picture}(300,300)(0,0)
\Gluon(20,160)(75,160){5}{4} \Vertex(75,160){2}
\GlueArc(150,160)(75,0,180){5}{16}
\Gluon(75,160)(135,160){5}{4} 
\DashLine(135,160)(165,160){6}
\Gluon(165,160)(225,160){5}{4} \Vertex(225,160){2}
\Gluon(225,160)(280,160){5}{4}
\Gluon(150,140)(150,85){5}{4}
\GCirc(135,160){4}{0.8}
\GCirc(165,160){4}{0.8}
\GCirc(150,140){4}{0.8}
\Text(90,35)[]{(a)}
\end{picture}}

\newcommand{\anfour}{\begin{picture}(300,300)(0,0)
\Gluon(50,280)(150,230){5}{5} 
\GlueArc(150,195)(35,-60,90){5}{6}
\DashLine(135,160)(165,160){6}
\Vertex(150,230){2}
\GlueArc(150,195)(35,90,240){5}{6}
\Gluon(150,230)(250,280){5}{5} 
\Gluon(150,140)(150,85){5}{4}
\GCirc(135,160){4}{0.8}
\GCirc(165,160){4}{0.8}
\GCirc(150,140){4}{0.8}
\Text(90,35)[]{(b)}
\end{picture}}

\newcommand{\brthree}{\begin{picture}(300,300)(0,0)
\Gluon(50,160)(135,160){5}{7}
\Gluon(165,160)(225,160){5}{5}
\DashLine(135,160)(165,160){6}
\Gluon(150,140)(150,85){5}{3}
\GlueArc(150,160)(75,-90,0){5}{10}
\GCirc(135,160){4}{0.8}
\GCirc(165,160){4}{0.8}
\GCirc(150,140){4}{0.8}
\Vertex(225,160){2}
\Vertex(150,85){2}
\Gluon(225,160)(285,160){5}{5}
\Gluon(150,35)(150,85){5}{3}
\Text(90,10)[]{(c)}
\end{picture}}

\newcommand{\brfour}{\begin{picture}(300,300)(0,0)
\Gluon(50,160)(130,160){5}{7}
\DashLine(135,160)(165,160){6}
\GlueArc(170,140)(20,-180,-45){5}{3}
\GlueArc(170,140)(20,-45,90){5}{3}
\GCirc(130,160){4}{0.8}
\GCirc(170,160){4}{0.8}
\GCirc(150,140){4}{0.8}
\Vertex(184.14,125.85){2}
\Gluon(184.14,125.85)(184.14,65.85){5}{5}
\Gluon(184.14,125.85)(244.14,125.85){5}{3}
\Text(90,10)[]{(d)}
\end{picture} }

\def\Am#1#2#3{\widetilde A_{#1}^{#2}(#3)}
\def\A#1#2#3{A_{#1}^{#2}(#3)}
\def\C2#1#2{({\cal C}_2)_{#1}^{#2}}

\def\eq#1{Eq. (\ref{#1})}

\def\jhep#1#2#3{J. High Energy Phys. {\bf #1} (#2) #3}
\def\prd#1#2#3{Phys.\ Rev.\ {\bf D#1} (#2) #3}
\def\npb#1#2#3{Nucl.\ Phys.\ {\bf B#1} (#2) #3}
\def\plb#1#2#3{Phys.\ Lett.\ {\bf B#1} (#2) #3}

\def\np#1#2#3{Nucl.\ Phys.\ B#1 (19#3) #2}

\def\zpc#1#2#3{Z.\ Phys.\ {\bf C#1} (#2) #3}
\begin{document}
%\today
\setcounter{page}{1}
\begin{flushright}
LPT-ORSAY 01-40\\
UHU-FT/01-02\\
\end{flushright}
%filename="puisalpha3.tex"
\begin{center}
\bf{\huge 
Remarks on \\ 
the determination of the Landau gauge OPE \\
for  the Asymmetric three gluon vertex}
\end{center}  
\vskip 0.8cm
\begin{center}
{\bf  F. De Soto$^{a}$, J. Rodr\'{\i}guez-Quintero$^{b}$   
}\\
%{\bf by US} \\
\vskip 0.5cm 
$^a${\sl Dpto. de F\'{\i}sica At\'omica, Molecular y Nuclear \\
Universidad de Sevilla, Apdo. 1065, 41080 Sevilla, Spain} \\
$^b${\sl Dpto. de F\'{\i}sica Aplicada e Ingenier\'{\i}a el\'ectrica \\
E.P.S. La R\'abida, Universidad de Huelva, 21819 Palos de la fra., Spain} \\
\end{center}
\begin{abstract}
\medskip

We compute a compact OPE formula describing  power corrections to the perturbative expression for
the asymmetric  $\MOM$-renormalized running coupling constant  up to the leading logarithm. By the 
use of the phenomenological hypothesis leading to the factorization of  the condensates through a
perturbative vacuum  insertion, the only relevant condensate in the game is $\langle A^2 \rangle$.
The validity  of the OPE formula is tested by searching for a good-quality coherent description of
previous  lattice evaluations of $\MOM$-renormalized gluon propagator and running coupling. 

\bigskip

\noindent P.A.C.S.: 12.38.Aw; 12.38.Gc; 12.38.Cy; 11.15.H

\end{abstract}

That the running coupling constant can be extracted from the three-gluon vertex in the Landau  gauge
was proposed several years ago in a seminal work \cite{Parri}. This method appears closer to a
physical interpretation than Scr\"odinger functional's \cite{alpha} but, and mainly for the same
reason, more  systematic effects should be managed to produce a reliable prediction for the
perturbative $\alpha_S$. In particular, the first statistically  meaningful attempts to follow that
method missed the impact of power  corrections\cite{frenchalpha} and failed to give an estimate for
the coupling constant comparable  to others in literature (that of ref. \cite{alpha}, for instance).
In parallel, these power  corrections to the gluon propagator and to the running coupling constant
have been largely  studied in the last years\cite{renormalons,lavelle}. 

Although the first trials were rather inconclusive\cite{Parri2}, momentum power  contributions have
been manifestly put in evidence\cite{poweral,OPE} for the Landau gauge three-gluon coupling constant
renormalized in both symmetric (MOM) and asymmetric ($\MOM$) momentum substraction schemes. In ref.
\cite{poweral} the parameter $\Lams$ is estimated from the matching of $\MOM$ $\alpha_S$ lattice
results to a perturbative three-loop formula corrected by an unavoidable $1/p^2$-term. This naive
ansatz used in ref. \cite{poweral} seems to eliminate most of the systematic deviation from the 
three-gluon vertex estimate of the perturbative $\alpha_S$ and a precise prediction of $\Lams$
emerges in full agreement with that of ref. \cite{alpha}. Unfortunately, the errors quoted in this
work were clearly underestimated. In fact, the prediction of $\Lams$ is so sensitive to a
logarithmic dependence on the momentum scale of the coefficient of $1/p^2$ that it appears to range
over an interval of 40 MeV\cite{OPEOne}. 

On the other hand, in refs. \cite{OPE,OPEOne} a description  in terms of OPE for the
power corrections to MOM three-gluon $\alpha_S$ is successfully  tried consistently with
an analogous description for gluon propagator.  The OPE approach provides through SVZ
factorization\cite{weinberg} a perturbative  tool to obtain the leading logarithmic
dependence of non-trivial Wilson  coefficients. In particular, it can be applied to
compute the coefficient of $A^2$ in the  MOM case for Landau gauge three-gluon coupling
constant\cite{OPEOne}.  Obviously, the expectation value of $A^2$ in a non-perturbative
vacuum is also estimated  via this OPE approach. A non gauge invariant  condensate,
that, being in Landau gauge, is invariant under infinitesimal gauge  transformations and
thus connected in some way to the gauge invariant  $\langle A^2_{\rm min} \rangle$
defined in \cite{Zakh}, is then computed. 

The same approach for $\MOM$ coupling constant requires to use

\beq
T^*\left( \Am{\mu}{a}{-p} \Am{\nu}{b}{p}\right)&=& \nonumber \\
(c_0)^{a b}_{\mu \nu}(p) \ 1 
&+& (c_1)^{a b \mu'}_{\mu \nu a'}(p) : \A{\mu'}{a'}{0}: \ + \
(c_2)^{a b \mu' \nu'}_{\mu \nu a' b'}(p) \ 
:\A{\mu'}{a'}{0} \ \A{\nu'}{b'}{0}:  \nonumber \\
&+& (c_3)^{a b \mu' \nu' \rho'}_{\mu \nu a' b' c'}(p) \ 
:\A{\mu'}{a'}{0} \ \A{\nu'}{b'}{0} \A{\rho'}{c'}{0}: \ + \   
\dots \;\; ; \label{OPEfield1} 
\eeq

\noindent where the expansion to the three-gluon local operator is necessary
because the three-point Green function in $\MOM$ can be written as

\beq
\langle \ T^*\left( \Am{\mu}{a}{-p} \Am{\nu}{b}{p} \Am{\rho}{c}{0}\right) \ 
\rangle_{\rm NP} 
&\equiv& \nonumber \\
\langle 0 | T^*\left( \Am{\mu}{a}{-p} \Am{\nu}{b}{p}\right) | g_\rho^c \rangle_{\rm NP} &=&
(c_1)^{a b \mu'}_{\mu \nu a'} \ \langle 0 |:\A{\mu'}{a'}{0}:| g_\rho^c \rangle_{\rm NP}
\nonumber \\
&+& (c_3)^{a b \mu' \nu' \rho'}_{\mu \nu a' b' c'} \ 
\langle 0 |:\A{\mu'}{a'}{0} \A{\nu'}{b'}{0} \A{\rho'}{c'}{0}:| g_\rho^c \rangle_{\rm NP} 
\ + \ \cdots \; \; .
\label{OPEGreen}
\eeq

\noindent The index NP refers to the non-perturbative nature of the vacuum state in Eq. 
(\ref{OPEGreen}), while $T^*$ refers to the standard time ordered product in  momentum space. It
should be noticed that: i) No other local operators with the same dimension of $A^2$  are written in
Eq. (\ref{OPEfield1}) because, unlike the identity or  $A^2$ itself, they do  not generate non-null
vacuum expectation value (\textit{v.e.v.})\footnote{No Lorentz  invariant tensor with an odd number
of indices can be built without non-zero momenta.}  ii) Operators other than the three-gluon local
one for the same dimension,  as $\partial_\mu \A{\nu}{b}{0} \A{\rho}{c}{0}$, do not appear 
explicitly because they are phenomenologically supposed not to survive, as will be  argued below. The
identity  and $A^2$ clearly do not contribute to the matrix element considered in Eq.
(\ref{OPEGreen})

Following standard SVZ techniques to obtain the peturbative expansion of  OPE Wilson coefficients,
we compute the appropriated matrix  element of Eq. (\ref{OPEfield1})'s l.h.s. to the wanted order.
It is immediate that  taking the perturbative vacuum leads to

\beq
\langle 0 | T^*\left( \Am{\mu}{a}{-p} \Am{\nu}{b}{p} \Am{\rho}{c}{0}\right) | 0 \rangle  
\ = \ 
(c_1)^{a b \mu'}_{\mu \nu a'} \ \widetilde{G^{(2)}}^{a' c}_{\mu' \rho}(0) \ ,
\label{c1}
\eeq

\noindent where $c_1$ can be straightforwardly identified with the perturbative  expansion for Green
Function with an amputated soft-gluon leg,

\beq
\Gamma^{a b \mu'}_{\mu \nu a'}(p,-p,0) 
\equiv - 2 p^{\mu'}g^\bot_{\mu \nu}(p) f^{a b}_{\ \ a'} \ G^{(3)}_{\rm pert}(p^2) \ ,
\label{tree}
\eeq

\noindent which should be proportional to its Landau gauge tree-level tensor.  Beyond this purely
perturbative first contribution, the situation is a bit  more complicated. We consider now the
matrix element in the l.h.s. of  Eq. (\ref{OPEGreen}) between two external soft gluons and the
perturbative vacuum. Then,  we obtain up to the considered order in perturbation theory

\beq
&& \hspace{-0.5cm} \langle 0 | T^*\left( \Am{\mu}{a}{-p} \Am{\nu}{b}{p} \Am{\rho}{c}{0}\right) 
| g_{\lambda}^{l} g_{\sigma}^{s} \rangle_{\rm connected}  \nonumber \\
&&= 
(c_3)_{\mu \nu a' b' c'}^{a b \mu' \nu' \rho'}(p) \ 
\langle 0 | :\A{\mu'}{a'}{0} \A{\nu'}{b'}{0} \A{\rho'}{c'}{0}: 
| g_{\rho}^c g_{\lambda}^{l} g_{\sigma}^{s} 
\rangle \ .
\eeq

\noindent For the matrix element in the r.h.s. we have

\beq
&& \hspace {-0.5cm} \langle 0 | :\A{\mu'}{a'}{0} \A{\nu'}{b'}{0} \A{\rho'}{c'}{0}: 
| g_{\rho}^c g_{\lambda}^{l} g_{\sigma}^{s} 
\rangle \nonumber \\ &&= \ 
\widetilde{G^{(2)}}_{\sigma \sigma'}^{s s'}(0) \ \widetilde{G^{(2)}}_{\lambda \lambda'}^{l l'}(0)
\ \widetilde{G^{(2)}}_{\rho \tau}^{c t}(0) \ \Big\{ {\cal P}_{\sigma' \lambda' \tau}^{s' l' t} 
\ g_{\mu'}^{\ \sigma'} g_{\nu`}^{\ \lambda'} g_{\rho'}^{\ \tau} \delta^{a'}_{\ s'}
\delta^{b'}_{\ l'} \delta^{c'}_{\ t} 
\ + \ O(\alpha) \ \Big\} \ ,
\eeq

\noindent where ${\cal P}$ refers to all the possible permutation of the couples $(\sigma' s')$, 
$(\lambda' l')$, $(\tau t)$. The Wilson coefficient $c_3$ may thus be computed at tree-level  order
as:

\beq
{\cal P}_{\sigma' \lambda' \tau}^{s' l' t}\ 
(c_3)_{\mu \nu s' l' t}^{a b \sigma' \lambda' \tau}=
\frac{\langle \Am{\lambda}{l}{0} \Am{\mu}{a}{-p} \Am{\nu}{b}{p} \Am{\rho}{c}{0} \Am{\sigma}{s}{0}
 \rangle}
{\widetilde{G^{(2)}}_{\sigma \sigma'}^{s s'}(0) \ \widetilde{G^{(2)}}_{\lambda \lambda'}^{l l'}(0)
\ \widetilde{G^{(2)}}_{\rho \tau}^{c t}(0)} \ .
\label{Diag}
\eeq

\begin{figure}[hbt]
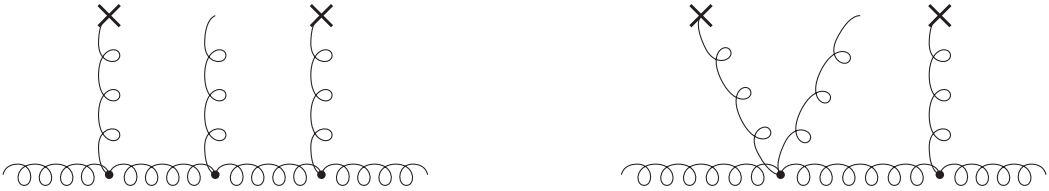

\vspace{30pt}
\begin{tabular}{cc}
\hspace*{-1cm} \SetScale{0.8} \vthree & 
\hspace*{-3cm} \SetScale{0.8} \vfour 
\end{tabular}
\vspace{-30pt}
\caption{\small Diagrams contributing to the tree-level Wilson coefficient in \eq{Diag}. Crosses
mark the soft gluon legs coming from the  condensate.}
\label{Fig1}
\end{figure}

\noindent The ratio in Eq. (\ref{Diag}) represents symbolically all the tree-level  diagrams with
five gluon legs where the three of them carrying zero momentum are cut (See fig. \ref{Fig1}).

\bigskip

Had we directly dealt with the three-gluon local operator  $:A^{a'}_{\mu'}(0) A^{b'}_{\nu'}(0)
A^{c'}_{\rho'}(0):$, the evaluation of  the higher order tensors involved  would require a much more
tedious calculation. On the other hand, the {\it vacuum insertion} between one gluon field and the
other two leads to 

\beq
\left[ \langle 0 |:A^{a'}_{\mu'}(0) A^{b'}_{\nu'}(0) 
A^{c'}_{\rho'}(0):| g_\rho^c \rangle_{\rm NP} \right]_{{\rm R}, \mu}
& = &  
\frac {\langle A^2 \rangle_{{\rm R}, \mu} \ \widetilde{G}_{\tau \rho}^{t c}(0,\mu^2)}
{ 4(N_c^2-1)} \ {\cal T}_{\mu' \nu' \rho' t}^{a' b' c' \tau}
\label{VI}
\eeq

\noindent with

\beq
{\cal T}_{\mu' \nu' \rho' t}^{a' b' c' \tau} \ = \
g_{\mu' \nu'}\delta^{a' b'}
g_{\rho'}^{\ \tau}\delta^{c'}_{\ t}+g_{\mu' \rho'}\delta^{a' c'}g_{\tau}^{\ \nu'}
\delta_{t}^{\ b'} + g_{\rho' \nu'}\delta^{c' b'}g_{\mu'}^{\ \tau}\delta^{a'}_{\ t} \ .
\eeq

\noindent We may phenomenologically assume vacuum  insertion to work for a certain renormalization
momentum scale.  Thus, the following replacement 

\beq
:A^{a'}_{\mu'}(0) A^{b'}_{\nu'}(0) A^{c'}_{\rho'}(0): \ \ \ \to \ \ \  
\frac {:A^2 A^{t}_{\tau}(0):}{ 4(N_c^2-1)} \ {\cal T}_{\mu' \nu' \rho' t}^{a' b' c' \tau}
\label{replace}
\eeq

\noindent can be done for Eq. (\ref{OPEGreen}), the other   components of the tensor in l.h.s. not
being required for our purposes.  Furthermore the local gluon field $\A{\tau}{t}{0}$ is to be
contracted with the  zero-momentum gluon field defined  for the vertex. The same vacuum insertion
assumption leads us to argue that, for instance,

\beq
\left[ \langle 0 |:\partial_\mu' A^{b'}_{\nu'}(0) 
A^{c'}_{\rho'}(0):| g_\rho^c \rangle_{\rm NP} \right]_{{\rm R}, \mu}
& \to &  
\langle \partial_\mu'  A^{b'}_{\nu'}(0) \rangle_{{\rm R}, \mu} \ 
\widetilde{G}_{\rho' \rho}^{c' c}(0,\mu^2) \ = \ 0 \ .
\label{Example}
\eeq 

\noindent Then, the only non-zero surviving condensate comes  from the three-gluon local operator.

It is easy to see that, using \eq{replace}, the relevant coefficient multiplying the  local operator
in Eq. (\ref{OPEfield1}) is 

\beq
(c_3)_{\mu \nu a' b' c'}^{a b \mu' \nu' \rho'}(p) \ {\cal T}_{\mu' \nu' \rho' t}^{a' b' c' \tau} 
\ = \
\frac{1}{2}\
\frac{\langle \Am{\lambda}{l}{0} \Am{\mu}{a}{-p} \Am{\nu}{b}{p} \Am{\rho}{c}{0} \Am{\sigma}{s}{0}
 \rangle}
{\widetilde{G^{(2)}}_{\sigma \mu'}^{s a'}(0) \ \widetilde{G^{(2)}}_{\lambda \nu'}^{l b'}(0)
\ \widetilde{G^{(2)}}_{\rho \tau}^{c t}(0)}\ g_{\mu' \nu'}\delta^{a' b'} \,
\label{Wilcoef}
\eeq

\noindent where the r.h.s. may be straightforwardly obtained from Eq. (\ref{Diag}).   This last
\eq{Wilcoef} gives the prescription in which Lorentz and color indices of  external gluon legs are
contracted in the diagrams contributing to the tree-level Wilson  coefficient (see Fig. \ref{Fig1}).

Thus, since Eq. (\ref{tree})'s is the only Landau gauge tensor for the asymmetric three-gluon 
vertex\cite{frenchalpha}, we can write for $p^2=-k^2$

\beq
&& \hspace{-0.5cm} k^4 \langle \ T^*\left( \Am{\mu}{a}{-p} \Am{\nu}{b}{p} \Am{\rho}{c}{0}\right) \ 
\rangle_{\rm NP}
\ = \ - 2 p^{\tau}g^\bot_{\mu \nu}(p) f^{a b}_{\ \ t} \nonumber \\
&& \times \ \left( c_1\left(g,\frac{k^2}{\Lambda^2}\right) \ 
\widetilde{G^{(2)}}^{t c}_{\tau \rho}(0) 
\ + \ 
c_3\left(g,\frac{k^2}{\Lambda^2}\right) \ 
\frac {\langle 0 |:A^2 A^{t}_{\tau}(0):| g_\rho^c \rangle_{\rm NP}}{ 4(N_c^2-1)} \
\frac{1}{-k^2} \ \right) \ . 
\label{bare}
\eeq

\noindent We know the scalar coefficients $c_1,c_3$ in Eq. (\ref{bare}) to be dimensionless by OPE
power counting rules, and hence both only depend on the bare  coupling $g$ and the dimensionless
ratio of momentum over  regularization scale~\footnote{The dependence on regularization momentum
scale  ($a^{-1}$ in lattice regularization or $\varepsilon^{-1} \mu$ in dimensional, for instance)
has  been up to now omitted to simplify the notation}.  The term $c_1$ can be obtained  from \eq{c1}
and, at tree-level, $c_3$ is to  be computed by projecting over the  Landau gauge tree-level tensor
in \eq{tree} the result from  \eq{Wilcoef},

\beq
c_{3,tree-level}\left(g,\frac{k^2}{\Lambda^2}\right) 
\ = \ (c_3)_{\mu \nu a' b' c'}^{a b \mu' \nu' \rho'}(p) \ 
{\cal T}_{\mu' \nu' \rho' t}^{a' b' c' \tau} \ 
\frac{1}{-6 N_C(N_C-1) k^2} p_\tau g^{\bot \ \mu \nu}(p) f_{a b}^{\ \ t} \nonumber \\ 
\ = \ 3 g . 
\label{Proj}
\eeq

\noindent If we renormalize following the $\MOM$ prescription at momentum scale  $\mu^2$, and then
apply the assumed vacuum insertion factorization,  we can write:

\beq
&& \hspace{-0.5cm} k^4 \left[ 
\langle \ T^*\left( \Am{\mu}{a}{-p} \Am{\nu}{b}{p} \Am{\rho}{c}{0}\right) \ 
\rangle \right]_{{\rm R},\mu^2}
\ = \ - 2 p^{\tau}g^\bot_{\mu \nu}(p) f^{a b}_{\ \ t} \
\widetilde{G^{(2)}}^{t c}_{\tau \rho}(0,\mu^2)
\nonumber \\ 
&& \times \ \left( c_1\left(\frac{k^2}{\mu^2},\alpha(\mu)\right) \ 
\ + \ 
c_3\left(\frac{k^2}{\mu^2},\alpha(\mu)\right) \ 
\frac {\langle A^2 \rangle_{\rm R},\mu^2}{ 4(N_c^2-1)} \
\frac{1}{-k^2} \ \right) \ . 
\label{Ren}
\eeq

\noindent where, after renormalization, Wilson coefficients depend on the ratio  of the momentum and
the renormalization scale, and on $\alpha(\mu)=g^2_R(\mu)/(4\pi)$, {\it i.e.} the  coupling constant
consistently renormalized in $\MOM$.

In the $\MOM$ scheme, renormalized Green functions take formally, at the  renormalization scale, the
same tree-level value but in terms of the renormalized  coupling constant instead of the bare one. 
This defines $Z^{\rm MOM}(\mu)=\mu^2 G^{(2)}(\mu^2)$ to renormalize appropriately the two-point
Green function, where $G^{(2)}$ is the scalar factor of the bare  two-point Green function defined
in ref. \cite{frenchalpha}. The three-gluon Green function  is then renormalized dividing by
$(Z^{\rm MOM}(\mu))^{3/2}$. One gets          

\beq
g_R(k^2)=k^4 \ \frac{G_R^{(3)}(k^2,\mu^2)}{G_R^{(2)}(0,\mu^2)} \
\left( k^2 G_R^{(2)}(k^2,\mu^2) \right)^{-1/2} \ .
\label{gR}
\eeq

\noindent The renormalized scalar factor for the three-gluon Green function with an amputated
soft-gluon, $G_R^{(3)}/G_R^{(2)}$,  can be projected out from \eq{Ren}, similarly to what is  done
for $c_3$ in \eq{Proj}, while  for $G_R^{(2)}(k^2,\mu^2)$ we can write 

\beq
k^2 \ G_R^{(2)}(k^2,\mu^2)=c_0\left(\frac{k^2}{\mu^2},\alpha(\mu)\right) \ + \
c_2\left(\frac{k^2}{\mu^2},\alpha(\mu)\right) \ 
\frac {\langle A^2 \rangle_{\rm R},\mu^2}{ 4(N_c^2-1)} \
\frac{1}{-k^2} \ ,
\label{G2}
\eeq

\noindent where the scalar coefficients $c_0$ and $c_2$ can be derived from those in \eq{OPEfield1}
and computed similarly to $c_1$ and $c_3$, as explained  in ref \cite{OPEOne}. Thus, after replacing
in \eq{gR} we will finally get:

\beq
&& \hspace{-0.5cm} g_R(k^2)=c_1\left(\frac{k^2}{\mu^2},\alpha(\mu)\right) \ 
\left[ c_0\left(\frac{k^2}{\mu^2},\alpha(\mu)\right) \right]^{-1/2} \nonumber \\
&& \times \ \left( \ 1 \ + \ \left( \frac{c_3\left(\frac{k^2}{\mu^2},\alpha(\mu)\right)}
{c_1\left(\frac{k^2}{\mu^2},\alpha(\mu)\right)} \ - \ \frac{1}{2} \ 
\frac{c_2\left(\frac{k^2}{\mu^2},\alpha(\mu)\right)}
{c_0\left(\frac{k^2}{\mu^2},\alpha(\mu)\right)} \right) \ 
\frac {\langle A^2 \rangle_{\rm R},\mu^2}{ 4(N_c^2-1)} \
\frac{1}{-k^2} \ \right) \ .
\label{gRf}
\eeq

\begin{figure}[hbt]
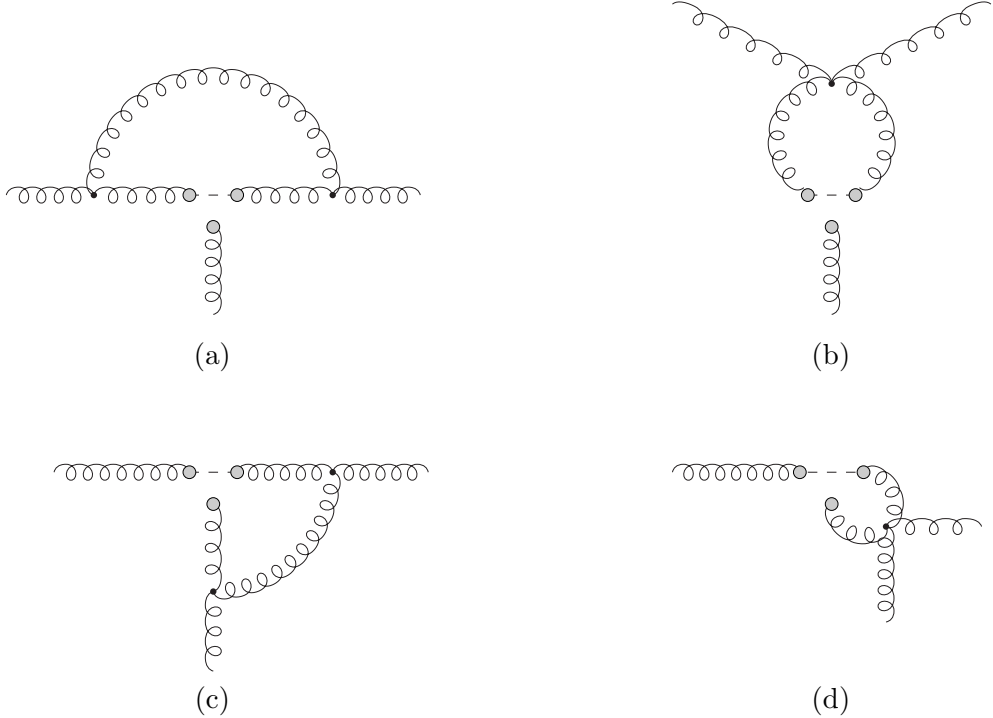

\vspace*{-4cm}
\begin{tabular}{cc}
\SetScale{0.6} \anthree & 
\hspace*{-3cm} \SetScale{0.6} \anfour \vspace*{-7cm} \\
\SetScale{0.6} \brthree & 
\hspace*{-3cm} \SetScale{0.6} \brfour \\
\end{tabular}
\caption{\small All the possible leg permutations from diagrams in the figure contribute to the
renormalization constant $\overline{Z}$ defined in the  text. (a) and (b)-like diagrams, which do
not break the assumed-to-work  factorization hypothesis, are renormalized by $\widehat{Z}$
previously  computed in \cite{OPEOne}. (c) and (d) breaking factorization diagrams give  $Z_\kappa$.
The local operators are drawn as gray bullets, the two of them  joined by a dashed line represent
the ones contracted to give $A^2$ in  Eqs. (\ref{Oper}-\ref{ZA3Oper}) through the replacement in
\eq{replace}.}
\label{Fig2}
\end{figure}

The purpose is now to compute to leading logarithms the subleading Wilson coefficients in  \eq{Ren},
as done in ref. \cite{OPEOne}. It will be, to  this goal, useful to consider the following operator
expansion, 

\beq
\frac{2}{9 k^2} \ p_{\tau}{g^\bot}^{\mu \nu}(p) f_{a b}^{\ \ t} \ 
\left[ T^*\left( \Am{\mu}{a}{-p} \Am{\nu}{b}{p} \Am{\rho}{c}{0}\right) \ 
\right]_{{\rm R},\mu^2} \nonumber \\ =  
\frac{c_3\left(\frac{k^2}{\mu^2},\alpha(\mu)\right)}{-k^6}
\left[ :A^2 A^{t}_{\tau}(0): \Am{\rho}{c}{0} 
\right]_{{\rm R}, \mu} \ + \ \dots \ \ ;
\label{Oper}
\eeq

\noindent where dots refer to terms with powers of $1/k$ other than 6.  If the vacuum expectation in
the non-perturbative vacuum is considered for  the r.h.s. of \eq{Oper}, the result under the vacuum
insertion hypothesis  is known to be diagonal in color and Lorentz spaces. Then we will contract in
both sides of \eq{Oper} with $g^{\rho \tau} \delta_{c t}$ and we will take the following matrix
element (see ref. \cite{Rafael})

\beq
\frac{2}{9 k^2} \ p^{\rho}{g^\bot}^{\mu \nu}(p) f_{a b c} \ 
\langle 0 | T^*\left( \Am{\mu}{a}{-p} \Am{\nu}{b}{p} \Am{\rho}{c}{0}\right)  
| g_\sigma^s  g_\lambda^l \rangle_{{\rm R},\mu^2} \nonumber \\ = \   
\frac{c_3\left(\frac{k^2}{\mu^2},\alpha(\mu)\right)}{-k^6} \ 
\langle 0 | :A^2 A_{c}^{\rho}(0): 
| g_\rho^c g_\sigma^s  g_\lambda^l \rangle_{{\rm R},\mu^2} \ + \ \dots \ \ ;
\label{MaEl}
\eeq

\noindent where we take external gluons in a perturbative vacuum and  carrying soft momenta. From
\eq{MaEl} we get

\beq
-\frac{2}{9} k^4 \ \frac{p^{\rho}{g^\bot}^{\mu \nu}(p) f_{a b c} \ 
\langle 0 | T^*\left( \Am{\mu}{a}{-p} \Am{\nu}{b}{p} \Am{\rho}{c}{0}\right)  
| g_\sigma^s  g_\lambda^l \rangle}
{\langle 0 | :A^2 A_{c}^{\rho}(0): 
| g_\rho^c g_\sigma^s  g_\lambda^l \rangle} \nonumber \\ 
\ = \  Z_3(\mu^2) Z_{A^3}^{-1}(\mu^2) \
c_3\left(\frac{k^2}{\mu^2},\alpha(\mu)\right)  
\ = \ Z_3^{-1/2}(\mu^2)  \overline{Z}^{-1} \ 
c_3\left(\frac{k^2}{\mu^2},\alpha(\mu)\right)  
\label{ZA3}
\eeq

\noindent where, always in $\MOM$ prescription,  $\widetilde{A}_R=Z_3^{-1/2}\widetilde{A}$ and
$Z_{A^3}$ is defined such that

\beq
\left[ :A^2 A_{c}^{\rho}(0): 
\Am{\rho}{c}{0} \right]_{{\rm R},\mu^2} \ = \
Z_{A^3}^{-1}(\mu^2) Z_3^{-1/2}(\mu^2) \  :A^2 A_{c}^{\rho}(0): 
\Am{\rho}{c}{0} \ \ ;
\label{ZA3Oper}
\eeq

\noindent while $\overline{Z}\equiv Z_{A^3} Z_3^{-3/2}$ is an useful notation for the  constant
renormalizing the matrix element for the three-gluon local operator where the external soft gluons
are explicitly cut. If we recover the divergent factor  $\widehat{Z}\equiv Z_{A^2} Z_3^{-1}$
introduced in ref. \cite{OPEOne} for the  matrix element of two-gluon local operator coming from
proper vertex corrections,  $\overline{Z}$ can be thought to be decomposed as

\beq
\overline{Z}(\mu^2) \ \equiv \ \widehat{Z}(\mu^2) \ Z_\kappa(\mu^2) \ ,
\label{comp}
\eeq

\noindent where $\widehat{Z}$ takes the divergent part coming from the diagrams for  the matrix
element in r.h.s. of \eq{MaEl} which can be factorized (diagrams (a,b)  of fig. \ref{Fig2}) as
in \eq{VI}, and  $Z_\kappa$ should be computed from those which can not (diagrams (c,d)  of
fig. \ref{Fig2}). 

\bigskip

Then, taking the logarithmic derivatives with respect to $\mu$ in both sides of \eq{ZA3}, we get the
following renormalization group differential equation:

\beq
\left\{ \ -2 \overline{\gamma}\left(\alpha(\mu)\right) -  \gamma\left(\alpha(\mu)\right)
+\frac{\partial}{\partial\ln{\mu}} +\beta\left(\alpha(\mu)\right)  
\frac{\partial}{\partial\alpha} \ \right\} \ c_3\left(\frac{k^2}{\mu^2},\alpha(\mu)\right) 
\ = \ 0 \ ;
\label{RG1}
\eeq

\noindent with the formal solution (see ref. \cite{Rafael})

\beq
c_3\left(\frac{k^2}{\mu^2},\alpha(\mu)\right) = c_3\left(1,\alpha(k)\right) \ 
\left( \frac{\alpha(k)}{\alpha(\mu)} 
\right)^{-\frac{2 \overline{\gamma}_0+ \gamma_0}{2 \beta_0}} \ ,
\label{ForSol}
\eeq

\noindent where

\beq
&& \overline{\gamma}\left(\alpha(\mu)\right) \ = \ 
\frac{d}{d\ln{\mu^2}} \ln{\overline{Z}(\mu^2)} \ = 
- \overline{\gamma}_0 \ \frac{\alpha(\mu)}{4 \pi} 
+ ... \ , \nonumber \\ 
&& \gamma\left(\alpha(\mu)\right) \ = \ \frac{d}{d\ln{\mu^2}} \ln{Z_3(\mu^2)} \ = 
- \ \left(
\gamma_0 \frac{\alpha(\mu)}{4 \pi} \ + \ \gamma_1 \left(\frac{\alpha(\mu)}{4 \pi}\right)^2
\ + \ \gamma_2 \left(\frac{\alpha(\mu)}{4 \pi}\right)^3 \ + \ \dots \ \right) \ , 
\nonumber \\
&& \beta\left(\alpha(\mu)\right) \ = \
\frac{d}{d\ln{\mu}} \ \alpha(\mu) \ = \ 
- \left( \frac{\beta_0}{2 \pi} 
\alpha^2(\mu) \ + \ \frac{\beta_1}{4 \pi^2} \ \alpha^3(\mu)
\ + \ \frac{\beta_2}{(4 \pi)^3} \ \alpha^4(\mu) \ 
+ \ \dots \ \right) \ .
\label{Ad}
\eeq

\noindent The boundary condition of \eq{RG1} is given by our $\MOM$-like prescription for the
renormalization  of the condensate by $Z_{A^3}$: the condensate is renormalized such that the
Wilson  coefficient takes the tree-level form at the renormalization point. Thus the prefactor
$c_3(1,\alpha(k))$ has to be matched at tree-level to \eq{Proj}, and the only solution  to the
leading logarithm is

\beq
c_3(1,\alpha(k))= 3 \left(g_R(k)\right)^3 \ 
\left[1+{\cal O}\left( \frac{1}{\log{(k/\Lambda_{\rm QCD})}} \right) \right].
\label{BC}
\eeq

\noindent Identifying the non-power-corrected term of \eq{gRf} to  the  purely perturbative coupling
constant,  $c_1\left(\frac{k^2}{\mu^2},\alpha(\mu)\right)$ is known to be

\beq
c_1\left(\frac{k^2}{\mu^2},\alpha(\mu)\right) \ = \
g_{\rm R,pert}(k^2) \ \left[ c_0\left(\frac{k^2}{\mu^2},\alpha(\mu)\right) \right]^{1/2} \ .
\eeq

\noindent Thus, we take from ref. \cite{OPEOne} the following leading order results:

\beq
c_0\left(\frac{k^2}{\mu^2},\alpha(\mu)\right) \ = \ 
\left(\frac{\alpha(k)}{\alpha(\mu)}\right)^{\frac{\gamma_0}{\beta_0}} \nonumber \\
c_2\left(\frac{k^2}{\mu^2},\alpha(\mu)\right) \ = \ 3 g^2_R(k^2)  
\left(\frac{\alpha(k)}{\alpha(\mu)}\right)^{-\frac{\widehat{\gamma_0}}{\beta_0}} \ ,
\label{Bd}
\eeq

\noindent where, as usual, $\widehat{\gamma_0}$ is defined by

\beq
\frac{d}{d\ln{\mu^2}} \ln{\widehat{Z}(\mu^2)} \ = 
- \widehat{\gamma}_0 \ \frac{\alpha(\mu)}{4 \pi} 
+ ... \ \ .
\eeq

\noindent \eq{gRf} can be then applied to obtain $\alpha_{\widetilde{\rm MOM}}=g^2_R/(4 \pi)$ 
which, after the appropriate Wick rotation takes in Euclidean space the following form:

\beq
\alpha_{\widetilde{\rm MOM}}(k^2) \ = \ \alpha_{\rm pert}(k^2) \ \left\{ \ 1 \ + \ 
\frac{T(\mu)}{k^2} \left[\ln{\left(\frac{k}{\Lambda}\right)}\right]^{
\frac{\widehat{\gamma}_0+\gamma_0}{\beta_0}-1} \ \left( 2 \left[ 
\frac{\ln{(\frac{k}{\Lambda})}}{\ln{(\frac{\mu}{\Lambda})}} \right]^
{\frac{\kappa_0}{\beta_0}} \ - 1
\right) \right\}
\label{alphanp}
\eeq

\noindent with

\beq
T(\mu) \ = \ \frac{6 \pi^2}{\beta_0} \frac{\langle A^2 \rangle_{R,\mu}}{N_C^2-1}
\left[\ln{\left(\frac{\mu}{\Lambda}\right)} \right]^{-\frac{\widehat{\gamma}_0+
\gamma_0}{\beta_0}} \ ,
\label{Tmu}
\eeq

\noindent $\langle A^2 \rangle_{R,\mu}$ being now the Euclidean condensate; and

\beq
\kappa_0=\overline{\gamma}_0-\widehat{\gamma}_0 \ ,
\label{kappa}
\eeq

\noindent as immediately follows from \eq{comp} if $\kappa_0$ is  defined from

\beq
\frac{d}{d\ln{\mu^2}} \ln{\widehat{Z}_\kappa(\mu^2)} \ = 
- \kappa_0 \ \frac{\alpha(\mu)}{4 \pi} 
+ ... \ \ .
\eeq

\noindent The following perturbative coefficients for the flavourless case,

\beq
\beta_0 = 11,\ \ \beta_1 = 51,\ \ \gamma_0 = \frac{13}{2} \ ,
\eeq

\noindent are universal, while coefficients in the $\widetilde{\rm MOM}$ scheme, 
$\beta_2$ was computed in \cite{frenchalpha}, and $\gamma_1$ and $\gamma_2$ in 
\cite{propag}

\beq
\beta_2 \simeq 4824. \ ,\ \ \gamma_1 = \frac{29}{8} ,\ \ \gamma_2 = 960 \ .
\eeq

\noindent In a recent paper \cite{OPEOne} we computed $\widehat\gamma_0$, from  diagrams identical
to these in fig. \ref{Fig2} (a) and (b); $\kappa_0$, defined in \eq{kappa} can be computed from
diagrams (c) and (d) in fig. \ref{Fig2} to  give:

\beq
\widehat\gamma_0 = \frac{3N_C}{4}, \ \ \ \kappa_0 = -\frac{9 N_C}{136} \ .
\eeq

\noindent We have proceeded to evaluate the diagrams for $\kappa_0$ in total analogy to the 
calculation of those for $\widehat\gamma_0$. For instance, we turned the diagrams to be 
infrared-safe by considering a momentum flow incoming to the local operator. The  details of the
procedure can be found in ref. \cite{OPEOne}. It should be noticed that $\kappa_0 <<
\widehat\gamma_0$ and that,  in practice, $\kappa_0/\beta_0 \simeq 0$ works as a good approximation
to simplify \eq{alphanp}. In other words, the scheme given by vacuum insertion factorization
reveals  itself to be coherent: {\it leading logarithm corrections violating the factorization 
induce a very small running with renormalization scale for the factorized tree-level  Wilson
coefficient}. 

An important  point is nevertheless to prove the assumption to work in order to obtain  a good
estimate of the Wilson coefficient for the asymmetric three-gluon Green function.   To dig into this
question, we have performed the same combined fits shown in ref. \cite{OPEOne}  for two and three
points Green functions, at three loops for leading Wilson coefficients,  to match lattice data taken
from refs. \cite{frenchalpha,propag}. The Euclidean OPE formula for the two point Green  function
is, from ref. \cite{OPEOne},

\beq
Z^{\rm MOM}_{\rm Latt}(k^2,a) & = &
Z^{\rm MOM}_{\rm Latt}(\mu^2,a) \ 
c_0\left(\frac{k^2}{\mu^2},\alpha(\mu)\right) \nonumber \\
& \times & \left( 1 \ + \ R(\mu) \ 
\left( \ln{\frac{k}{\Lambda}}\right)^{\frac{\gamma_0 + 
\widehat{\gamma}_0}{\beta_0}-1} \ \frac{1}{k^2} \ \right) \; ;
\label{ZLatt}  
\eeq

\noindent where 

\beq
\frac{Z^{\rm MOM}_{\rm Latt}(k^2,a)}{Z^{\rm MOM}_{\rm Latt}(\mu^2,a)} \ = \ 
k^2 G^{(2)}_R(k^2,\mu^2) \ + \ O(a^2) \ ,
\label{ZLattInd}
\eeq

\noindent and

\beq
R(\mu)\ = \ \frac{6 \pi^2}{\beta_0 \ (N_c^2-1)} \ 
\left( \ln{\frac{\mu}{\Lambda}} 
\right)^{-\frac{\gamma_0 + 
\widehat{\gamma}_0}{\beta_0}} \langle A^2 \rangle_{{\rm R},\mu} \; .
\label{R2}
\eeq

\begin{figure}[hbt]
%\vspace*{-1.cm}
\hspace*{-1.3cm}
\begin{center}
\begin{tabular}{cc}
\epsfxsize6.8cm\epsffile{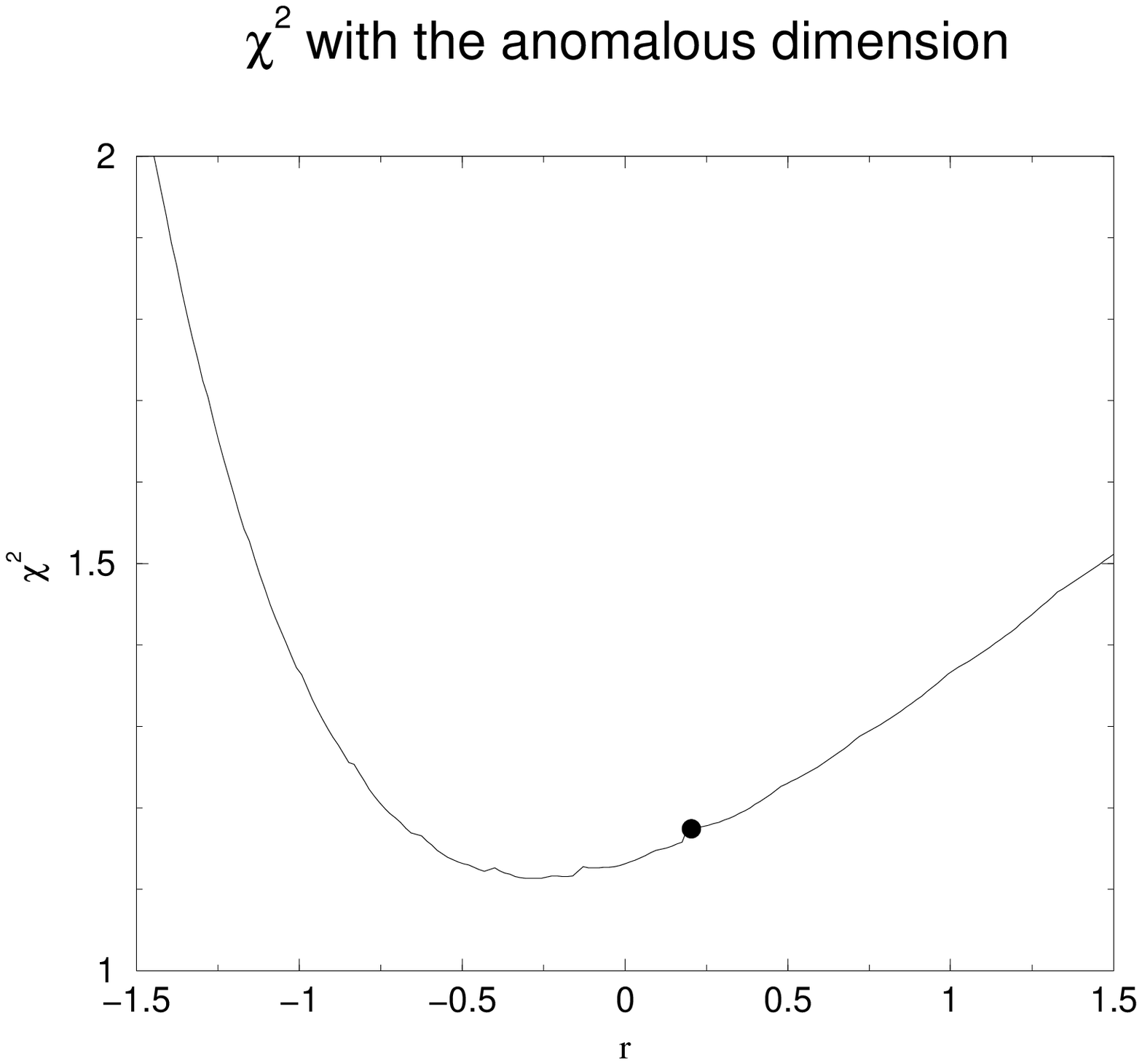} & 
\epsfxsize6.8cm\epsffile{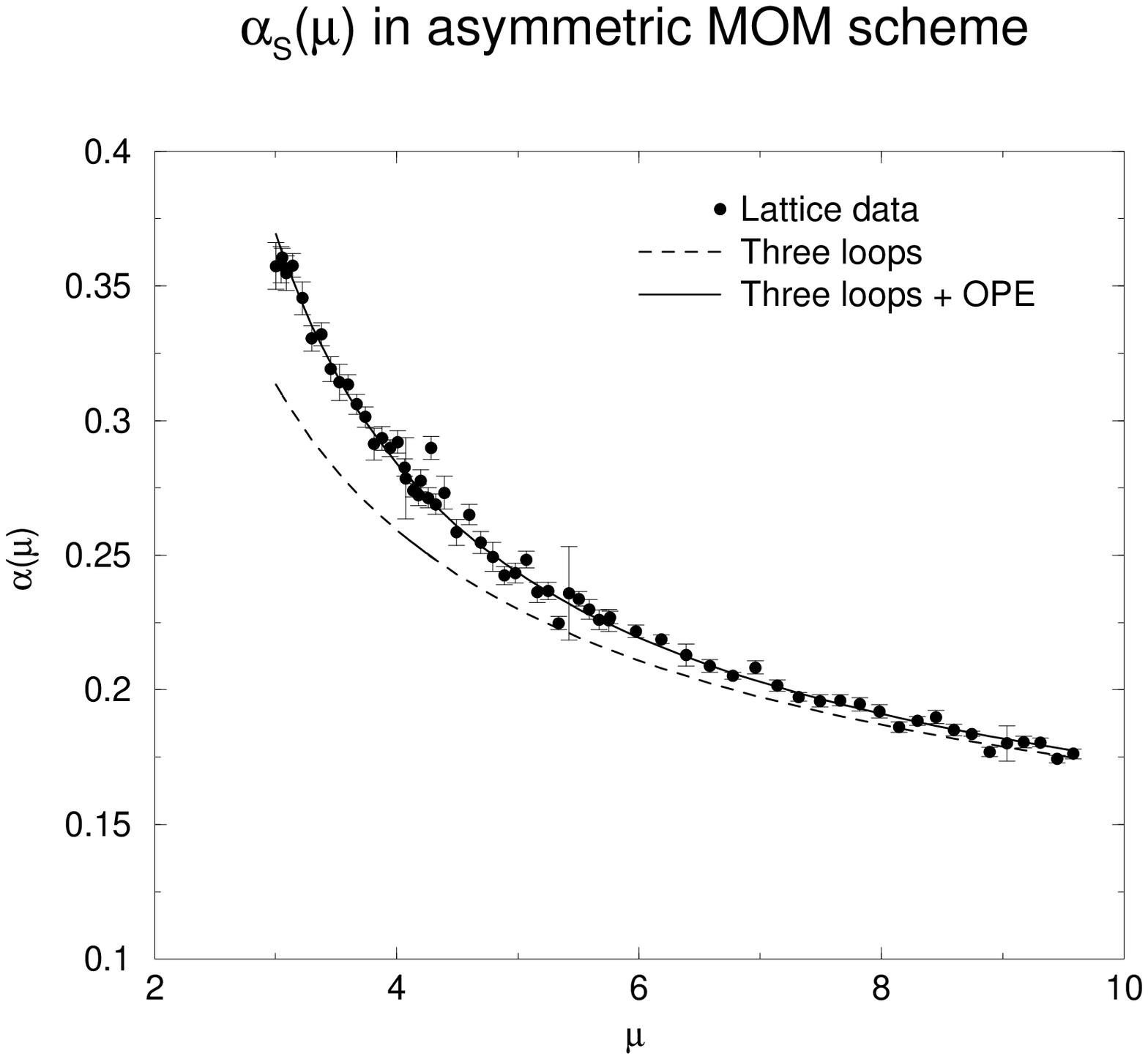} \\
(a) & (b) \\
\end{tabular}
\caption{\small {\it (a) Quality of the fit as a function of the exponent 
$r=1-\frac{\widehat{\gamma}_0+\gamma_0}{\beta_0}$ in Eq. \ref{alphanp}. The dot stands for  the
value of $r$ computed in this paper. In figure (b) is shown the fit of our  Lattice data for
$\alpha_{\widetilde{MOM}}$ at three loops with the calculated anomalous dimension.}}
\label{Fig3}
\end{center}
\end{figure}

\noindent The coefficient $c_0(\frac{k^2}{\mu^2},\alpha(\mu))$ is taken now to be  expanded  at
three loops in  terms of the $\MOM$ scheme $\alpha(k)$; {\it i.e.} it will verify the same 
differential  equation as $\gamma(\alpha)$ in \eq{Ad} with the boundary condition which is apparent
from \eq{Bd}. In the following, all the scale-dependent quantities will be shown at  $\mu=10$ GeV.
Furthermore, we have checked that both the ratio of  gluon condensate estimates and $\Lams$ indeed
do not depend on this last  momentum scale. The quality of the fits as a function of a free exponent
of  $\ln{\left(\frac{k}{\Lambda}\right)}$ in \eq{alphanp} has been explored and the   results are
shown in fig. \ref{Fig3}(a). We can conclude from these results that  the approach given by vacuum
insertion factorization provides a good estimate of this exponent. 

%%%%%%%%%%%%%%%%%%%%%%%%%%%%%%%%%%%%%%%%%%%%%%%%%%%%%%%%%%%%%%%%%%%%
%%%%%Figure 3
%%%%%%%%%%%%%%%%%%%%%%%%%%%%%%%%%%%%%%%%%%%%%%%%%%%%%%%%%%%%%%%%%%%%
%%%%%%%%%%%%%%%%%%%%%%%%%%%%%%%%%%%%%%%%%%%%%%%%%%%%%%%%%%%%%%%%%%%%%
%%%%%%%%%%%%%%%%%%%%%%%%%%%%%%%%%%%%%%%%%%%%%%%%%%%%%%%%%%%%%%%%%%%%%
\begin{table}[hbt]
\begin{center}
\begin{tabular}{||c||c|c||}
\hline 
\hline 
 & This work & symmetric 3-point \\
 \hline
$\Lambda_{\overline{\rm MS}}$ & 260(18) MeV & 233(28) MeV\\
\hline 
$\left\{\sqrt{\langle A^2 \rangle_{R,\mu}}\right\}_{prop} $ &  1.39(14) GeV & 1.55(17) GeV\\
\hline 
$\left\{\sqrt{\langle A^2 \rangle_{R,\mu}}\right\}_{alpha}$ & 2.3(6) GeV & 1.9(3)GeV \\
\hline 
\hline 
\end{tabular}
\caption{\small Comparison between results obtained for the three loops fit in the present work
 and in a previous one \cite{OPEOne}}
\label{tab1}
\end{center}
\end{table}
%%%%%%%%%%%%%%%%%%%%%%%%%%%%%%%%%%%%%%%%%%%%%%%%%%%%%%%%%%%%%%%%%%%%%%

The results for two particular values of the exponent $r$ in  Fig. \ref{Fig3}(a) are of interest for
the sake of comparison.  The case $r=-1$ corresponds to the formula proposed in ref.  \cite{poweral}
to be matched to the lattice data: a perturbative three-loop formula +  a term $c/p^2$, $c$ being a
constant. On the other hand, had we neglected the leading  logarithm contributions,
$\hat{\gamma}_0=\gamma_0=0$, the exponent would be $r=1$.  It can be seen from Fig. \ref{Fig3} that
both values of $r$ generate rather less good fits  to the lattice data.

Then, Eqs. (\ref{alphanp},\ref{Tmu}) can be used to perform fits at two and three loops for the
leading Wilson coefficients in order  to estimate, from asymmetric three-gluon Green function, the
gluon condensate.  The results of such a fits, plotted in fig. \ref{Fig3}(b), are: 

\beq
\frac{\left\{\sqrt{\langle A^2 \rangle_{R,\mu}}\right\}_{alpha}}
{\left\{\sqrt{\langle A^2 \rangle_{R,\mu}}\right\}_{prop}}= 3.65(4) \ \ \ 
\Lambda_{\overline{MS}} = 283(15) \Mev \ \ \ 
\chi^2 = 1.95
\label{rat2}
\eeq

\noindent for the two loops fit, and 

\beq
\frac{\left\{\sqrt{\langle A^2 \rangle_{R,\mu}}\right\}_{alpha}}
{\left\{\sqrt{\langle A^2 \rangle_{R,\mu}}\right\}_{prop}}= 1.7(3) \ \ \ 
\Lambda_{\overline{MS}} = 260(18) \Mev \ \ \ 
\chi^2 = 1.18
\label{rat3}
\eeq

\noindent for the three loops one. 

The impressive improvement from two to three loops suggests that the approach presented in this
work  permits a reasonable approximation to the Wilson coefficient.  The ratio decreases to almost
two $\sigma$'s away from $1$ and both estimates of $\Lambda_{\overline{\rm MS}}$ and of the  gluon
condensate turn to be close of the previous estimates obtained from the symmetric three-gluon Green
function in ref. \cite{OPEOne} (see table \ref{tab1}).  The scheme in this work and that of ref.
\cite{OPEOne}  differ only by the kinematics of the renormalization point. Such a different 
renormalization for Green functions implies a different renormalization of the gluon  condensate.
However, the discrepancy for estimates of $\langle A^2 \rangle_R$ in the two  works is expected not
to be important\footnote{To estimate the discrepancy  for the non-perturbative estimates of $\langle
A^2 \rangle_R$ we need to compute beyond leading  logarithm corrections that is out of the scope of
this work.}, as indeed can be seen in table  \ref{tab1}.  The comparison of these estimates, mainly
the ones obtained from gluon  propagator, strongly supports the claimed rather large contribution 
from the $A^2$ condensate\cite{poweral,OPE,OPEOne} that might be in connection with the  tachyonic
gluon mass scale studied in ref. \cite{Zakh}.

A negative hint regarding to previous results from the symmetric three-point Green  function is
nevertheless the higher central value of the ratio in \eq{rat3} ($1.2$ in  ref. {\cite{OPEOne}). In
principle two possible  sources of discrepancies could be expected: either  three loops is still
insufficiently accurate for the estimate of the perturbative part in the $\MOM$ renormalization
scheme, or there is a deviation from the assumed vacuum insertion (or factorization)  approximation.
Both effects would have a direct impact on the bigger ratio we obtain for asymmetric $\MOM$ scheme. 
The very good agreement between the gluon condensates estimated from  gluon propagator previously
discussed seems to point out the factorization breaking as the major contributing factor. Still the
ratio in Eq. (\ref{rat3}) is only about two sigmas from 1, which is in our opinion a rather
encouraging result. The $\langle A^2 \rangle$ deduced  from the propagator is in fair agreement with
previous estimates, even though it is biased by the factorization hypothesis through the fit of 
$\Lams$ which combines the propagator  and $\alpha_S$. This good result of the propagator   as well
as fig. \ref{Fig3}(a), suggests that our formula describing the power  corrections to $\alpha_S$ up
to the leading logarithm yields a good approximation of the exact one.

A two-sided goal is thus achieved: 

i) the results of ref. \cite{OPEOne} turn out to be confirmed by the use of a slightly  different
renormalization scheme.  

ii) The vacuum insertion factorization applied to condensates playing the game of the OPE for  the
asymmetric three-gluon Green function results in a compact prediction for its OPE power 
corrections. The coefficient of the power correction has been computed to the leading  logarithm,
and thus a most important source of systematic uncertainty for the estimate of  $\Lams$ in ref.
\cite{poweral} is eliminated. The latter is a positive feature because the $\beta$ function is
perturbatively known at four loops in the asymmetric $\MOM$~\cite{Chety2}  and lattice evaluations,
on the other hand, turn out to be statistically more precise in this  last renormalization scheme. 

A last consequence of this work and those from refs. \cite{OPE,OPEOne}: they althogether lead to
conclude that the Green functions methods, three-gluon vertex in particular,  provide us with a
reliable and precise enough estimate for the running coupling constant  and $\Lams$, once power
corrections are properly taken in consideration.

\section{Acknowledgments} We strongly acknowledge Ph. Boucaud, A. Le Yaouanc,  J.P. Leroy, J.
Micheli and O. P\`ene for the fruitful comments and stimulating  discussions held at the L.P.T. in
Orsay. The authors are also indebted to  J.A. Caballero and M. Lozano for reading carefully the
manuscript.  This work is supported in part by the European Community's Human Potential Program 
under contract HPRN-CT-2000-00145, Hadrons Lattice QCD, the spanish Fundaci\'on  C\'amara and
spanish DGICyT under contracts PB98-1111 and FPA2000-1592-C03-02.


\begin{thebibliography}{9}
{\small
\bibitem{Parri}
        B.~Alles, D.~Henty, H.~Panagopoulos, C.~Parrinello, C.~Pittori and D.~G.~Richards,
        \npb{502}{1997}{325}.
\bibitem{alpha}
        S. Capitani, M. Guagnelli, M. L\"uscher, S. Sint, R. Sommer, 
        P. Weisz and H. Wittig,
        Nucl. Phys. Proc. Suppl. {\bf 63} (1998) 153;
        Nucl. Phys. {\bf B544} (1999) 669.
\bibitem{frenchalpha}
        Ph. Boucaud, J. P. Leroy, J. Micheli, O. Pene, 
        C. Roiesnel, \jhep{10}{1998}{017}; \jhep{12}{1998}{004}.
\bibitem{renormalons}
        G Burgio, F. Di Renzo, G. Marchesini and E. Onofri, 
        \plb{422}{1998}{219}. \\ 
        For reviews and classic references see:\\
        V.I. Zakharov, \np{385}{1992}{452};\\
        A.H. Mueller, in {\it QCD 20 years later}, vol.~1 
        (World Scientific, Singapore 1993).
        B. Lautrup, \plb{69}{1977}{109};
        G. Parisi, \plb{76}{1977}{65}; \npb{150}{1979}{163};
        G. t'Hooft, in {\it The Whys of Subnuclear Physics}, Erice
        1977, ed A. Zichichi, (Plenum, New York 1977);
        M. Beneke and V.I. Zakharov, \plb{312}{1993}{340};
        M. Beneke \npb{307}{1993}{154};
        A. H. Mueller, \npb{250}{1985}{327}; \plb{308}{1993}{355};
        G. Grunberg, \plb{304}{1993}{183}; \plb{325}{1994}{441}.
\bibitem{lavelle} 
        M. Lavelle and M. Oleszczuk, Mod. Phys. Lett. { A 7} (1991)3617;
        J. Ahlbach, M. Lavelle, M. Schaden, A. Streibl, \plb{275}{1992}{124}. 
\bibitem{Parri2}
        G. Burgio, F. Di Renzo, C. Parrinello and C. Pittori
        Nucl. Phys. Proc. Suppl. 73 (1999) 623; Nucl. Phys. Proc. Suppl. 74 (1999) 388;
        hep-ph/9808258.
\bibitem{poweral}
        Ph. Boucaud {\it et al.}, \jhep{04}{2000}{006}. 
\bibitem{OPE}
        Ph. Boucaud, A. Le Yaouanc, J.P. Leroy, J. Micheli, 
        O. P\`ene, J. Rodriguez-Quintero, \plb{493}{2000}{315}.
\bibitem{OPEOne} 
        Ph. Boucaud,A. Le Yaouanc, J.P. Leroy, J. Micheli, 
        O. P\`ene, J. Rodriguez-Quintero, \prd{63}{2001}{114003} 
\bibitem{weinberg}              
        M.A. Shifman, A.I. Vainshtein, V.I. Zakharov, \npb{147}{1979}{385},447,519;
        M.A. Shifman, A.I. Vainshtein, M.B. Voloshin, V.I. Zakharov, \plb{77}{1978}{80};
        S. Weinberg, {\it The quantum theory of fields}, vol. 2 (Cambridge University Press
        1996)
\bibitem{Zakh} 
        F.V. Gubarev, V.I. Zakharov, hep-ph/0010096; 
        F.V. Gubarev, M.I. Polikarpov, V.I. Zakharov, hep-ph/9908292;
        K.G. Chetyrkin, S. Narison, V.I. Zakharov, \npb{550}{1999}{353}.
\bibitem{Rafael}
        P. Pascual, E. de Rafael, \zpc{12}{1982}{127}.
\bibitem{propag}
        D. Becirevic, Ph. Boucaud, J. P. Leroy, J. Micheli,
        O. Pene, J. Rodriguez-Quintero, C. Roiesnel, \prd{60}{1999}
        {094509}; \prd{61}{2000}{114508}.
\bibitem{Chety2}
        K.G. Chetyrkin, T. Seidensticker, hep-ph/0008094.
}
\end{thebibliography}
\end{document}